\documentclass[twocolumn,prb,showpacs, superscriptaddress]{revtex4}

\usepackage{graphicx}
\usepackage{dcolumn}
\usepackage{bm}
\usepackage{amssymb}
\usepackage{amsmath}

\begin{document}

\title{
Statistical properties of the critical eigenstates in power-law random banded 
matrices across the band
}

\author{C.~J.~ Paley}  
\email{cjp38@cam.ac.uk}  
\affiliation{Department of Chemistry, University of Cambridge,  
             Cambridge CB2 1EW, UK}   
\author{S.~N.~Taraskin}  
\affiliation{St. Catharine's College and Department of Chemistry, University of Cambridge,  
             Cambridge, UK}
\author{S.~R.~ Elliott}  
\affiliation{Department of Chemistry, University of Cambridge,  
             Cambridge CB2 1EW, UK}

\date{\today}

\begin{abstract}
The level-spacing distribution 
in the tails of the eigenvalue bands of the 
power-law random banded matrix (PRBM) ensemble have been investigated numerically.  
The change of level-spacing statistics 
across the band is examined for different coupling strengths and compared to the density of states for the different systems.  
It is confirmed that, by varying the eigenvalue region,
the same level-spacing statistics can be reached as by
varying the coupling strength.

\end{abstract}


\maketitle


The Anderson metal-insulator transition (MIT) \cite{Anderson_58} is 
a phenomenon of major physical importance that continues to attract a 
substantial research effort \cite{Brandes:book, Janssen:book}.  
With just short-range interactions, localization-delocalization (LD) 
transitions are only found in systems with dimensionality, $D$, greater than two \cite{Mirlin_00}.  
However, with the addition of long-range interactions, 
or correlations between the short-range interactions, it is possible 
to study the LD transition in systems with dimensionality less than two \cite{Levitov_90}.  
In this respect, 
power-law random-banded matrices (PRBMs), that exhibit this transition, 
have recently attracted much attention \cite{Cuevas_Euro, Lima_04, Garcia_04, Cuevas_02, Cuevas_01}.  
The PRBM ensemble was introduced by Mirlin et. al. \cite{Mirlin_96} 
and, in the real case, is defined as the ensemble of $N\times N$ random 
symmetric matrices, $\hat H$.  
The PRBM elements, $H_{ij}$, are randomly drawn from a Gaussian distribution, 
centred around zero, with a variance governed by a power-law decay: 
\begin{equation}
\langle |H_{ij}|^2 \rangle =\frac{|i-j|}{1+(\frac{|i-j|}{b})^{2\alpha}}~, 
\label{e1}
\end{equation}
where $\alpha$ and $b \in (0, \infty)$ are parameters.  Regarding $\hat H$ as a Hamiltonian, the eigenvalues, E, are energies.

For $\alpha=1$, it has been shown \cite{Mirlin_96} that all the eigenstates of these 
matrices are critical (i.e. at the LD transition).  
The parameter $b$ is inversely related to the coupling strength between the nodes.  
In the limit $b \gg 1$ and $\alpha=1$, the PRBM critical states are analogous to the 
critical states at the Anderson transition 
with $D=2+ \epsilon$ and $\epsilon \ll 1$, and   
for $b \ll 1$ to those 
found in the Anderson model with $D \gg 1$ \cite{Evers_00}.  
By varying $b$, it is possible to access a set of different critical 
theories  parameterized by dimension ($2 < D < \infty $) 
in the conventional Anderson transition.   
This ability to examine Anderson transitions in 
different dimensions, in the same effectively one-dimensional model, 
makes the study of PRBMs a powerful method to make progress 
in this rich field.  
As well as being an analogue for the study of important transitions elsewhere, 
the PRBM is physically important in its own right and has been applied 
to the study of the finite-temperature Luttinger liquid \cite{Kravtsov_00}, 
the coherent propagation of two interacting particles 
in a 1D weak random potential \cite{Ponomarev_97} 
and other problems \cite{Lima_04, Cuevas_04}.  
Recently, it has also been realized that, with the addition of 
chirality, the PRBM also describes the LD transition 
of quark zero modes in QCD \cite{Garcia_04}.  
In the QCD vacuum, the quark zero-mode wavefunction decay has a power-law dependence 
and long-range hopping between sites is possible.  
In  this model, the eigenvalues away from the centre of the spectral 
band are not affected by the chiral structure; this makes it relevant 
to the analysis of the critical states away from the band centre given 
in this paper.

It is known ~\cite{Evers_00} that, by changing the coupling parameter, $b$, for $\alpha=1$, it is 
possible to access the set of models in different dimensions by investigating 
the statistical properties of the PRBM eigenstates, $E$, 
around the band centre, 
$E=0$. 
Alternatively, this can be achieved by varying $E$ away from the 
band centre and keeping $b$ constant, as was mentioned in 
Ref.~\cite{Mirlin_00}. 
However, this feature of the PRBM has not been previously studied and 
confirmed numerically, and this is the aim of this paper:  
to investigate the statistical properties of the eigenstates of the PRBM 
across the whole range of the spectrum. 
One of the main results of the paper is to demonstrate 
numerically that there is a mapping between different values of $b$ and $E$
given certain critical level-spacing statistics,  i.e. 
$(|b'|,|E'|) \Leftrightarrow (|b''|,|E''|)$ with $|b''|>|b'|$ and $|E''|>|E'|$, 
in terms of equivalent level-spacing statistics. 
This is important, both in models where PRBMs are used to model systems 
in which all the eigenvalues are of physical importance, 
 see e.g. Ref. \cite{Garcia_04}, 
and to support the validity of taking statistics from a finite 
region ~\cite{Evers_00, Cuevas_01, Lima_04} around the band centre; we obtain estimates to justify the width 
of the region around the band centre that is used for statistical analysis.

\begin{figure}
\includegraphics[angle=270, width=0.48\textwidth]{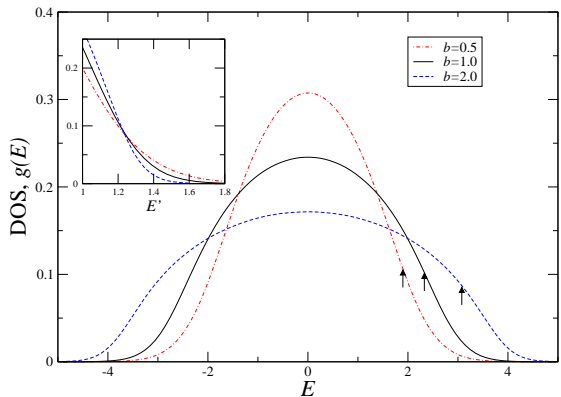}
\caption{\label{dos}Density of states of the PRBM ensemble with $\alpha=1$ 
and varying values of $b$ as marked.  
The inset shows portions of the densities of states after scaling so that 
90\% of the states are within the interval $E' \in [-1, 1]$.  
The arrows show the eigenvalue position corresponding to the 95th percentile
of the DOS.}
\end{figure}

To study the level-spacing statistics, 
a careful unfolding procedure was used, taking into account the 
varying mean level spacing (especially in the band tails) 
when normalizing the spacings. 
The width of the bins in $E$ was chosen to be sufficiently small so as to ensure that 
the nature of the states, and thus their statistical properties, do not change 
appreciably across them.  
The symmetry of the spectrum about $E=0$ has also been used to improve the
 level-spacing statistics. 
The size of the $N\times N$ matrices (with periodic boundary conditions) 
 varied in the range between  $N=1000$ and $N=8000$. 
The off-diagonal elements of the matrices were multiplied by a factor $0.5$ 
(as in Ref.~ \cite{Cuevas_03}) in order to 
 reinforce the closeness of the system to the Gaussian orthogonal ensemble.

The densities of states (DOS) of the PRBM ensembles for different values 
of the parameter $b$ and $\alpha=1$  are shown in Fig.~\ref{dos}.  
It can be seen that with increasing $b$ the bandwidth increases.  
There is substantial deviation from the semi-circle shape (cf. Ref.~ \cite{Varga_00})
 typical for the spectra of matrices from the Gaussian orthogonal 
ensemble, which do not show any change 
in the level-spacing statistics (the Wigner surmise) across the band \cite{Mehta:book}.  
The inset in Fig.~\ref{dos} shows eigenvalue-rescaled DOSs, 
and 
 verifies what is suggested by the main graph - 
that lower values of  $b$ also correspond to longer tails appearing in the density of states.  
This means that the DOSs corresponding to different values of $b$ cannot be equated  
by a simple linear rescaling of the eigenvalues. 
The fact that the plots in the inset appear to intersect at one point is not 
significant, but merely a result of the particular variables used to produce the graph. 

The bands shown in Fig.~\ref{dos} contain only critical states  
of multifractal nature, which exhibit a particular type of level-spacing 
statistics. 
For these critical states, 
the probability distribution function, $P_{\text{c}}(s)$, of normalized level spacings, $s$, 
between adjacent levels 
starts from zero at $s=0$ (level-repelling effect), goes through a maximum, 
and then exponentially decays for $s\gg 1$ according to the following law 
\cite{Zharekeshev_97, Bogomolny_99,Cuevas_Euro}:
\begin{equation} 
P_{\text{c}}(s) \propto  e^{-As^{\beta}}~.
\label{e1}    
\end{equation}
This functional form is different from those appropriate for descriptions of level-spacing 
statistics for localized or extended states. 
The level spacing for localized states follows the Poisson law, 
$P_{\text{P}}(s)=\exp{(-s)}$, while for extended states, the distribution 
is described by the  Wigner-Dyson surmise distribution ~\cite{Mehta:book}, 
$P_{\text{W}}(s)=(\pi/2) s \exp{(-(\pi/4) s^2)}$.  

In the case of the standard Anderson transition for systems in Euclidean geometries, 
there has been some controversy about the value of the exponent $\beta$ in Eq.~(\ref{e1}). 
Recent papers\cite{Varga_95, Cuevas_Euro} have given $\beta \simeq 1.2$ for 3D, 
whilst this value was refuted in another paper\cite{Zharekeshev_97} 
and the value set at $\beta=1.0 \pm 0.1$.
The uncertainty in the numerical value of $\beta$ for the standard Anderson transition comes 
from the fact that it is difficult  
to evaluate this exponent from an analysis of 
the critical states which exist only around one point (the LD transition) on the 
energy scale. 
This difficulty is lessened for the PRBM where all the states are critical, 
and thus an analysis of the level-spacing statistics can be performed relatively easily 
under the assumption that the statistics do not change with varying $E$. 
This assumption has been made~\cite{Cuevas_Euro} to perform a level-spacing 
statistical analysis around $E=0$. 
It was demonstrated there 
that the exponent $\beta$ in Eq.~(\ref{e1}) depends on the parameter $b$.  
An obvious question, which has not been addressed before, is how 
the level-spacing statistics change away from the band centre. 
In particular, is there a correspondence between the statistics for different values 
of $E$ and $b$? These questions are answered below. 

\begin{figure}
\includegraphics[angle=270, width=0.48\textwidth]{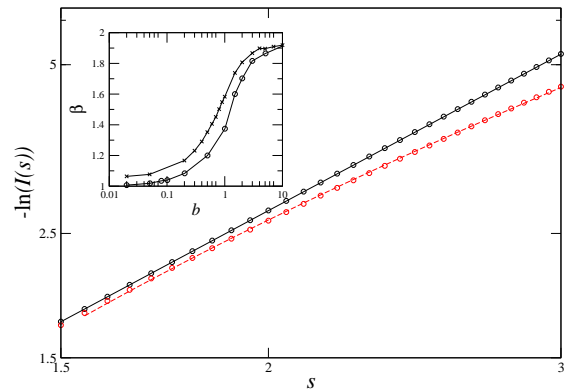}
\caption{\label{straight}
Double logarithmic plots of $-\ln{I(s)}$ versus $s$ for $\alpha=1$, $b=1$, 
in the central eigenvalue region $|E|\in [0,0.3]$ with a straight-line fit (solid line) to the 
displayed region, giving a slope $\beta=1.58$ and the band-tail region $|E|\in [1.6,1.7] $ (dashed line), yielding a 
gradient of 1.38.  
Configurational averaging over $10^5$ realizations 
of $1000\times 1000$ matrices 
was used for the statistical analysis. 
The inset shows how $\beta$ varies with $b$ at the 
band centre obtained by linear fits in the range $s \in [1.5, 3]$ (crosses)
with  the similar dependence obtained~\cite{Cuevas_Euro} by 
linear fits for $s>2$ (circles).
(The curves in the inset are guides to the eye.)
}
\end{figure}

In order to examine the level-spacing statistics for different values of $E$, 
we have performed a standard analysis, studying 
the cumulative level-spacing distribution 
$I(s)=\int_{s}^{\infty}P({s'})\text{d}{s'}$, 
which retains the asymptotic behaviour of $P(s)$.  
We expect a linear dependence in a double-logarithmic plot of $\ln(I)$ vs $s$ for $s\gg 1$, 
with the gradient (i.e. $\beta$) being in the range 
$\beta_{\text{P}}< \beta < \beta_{\text{W}}$, where 
$\beta_{\text{P}}=1$ and $\beta_{\text{W}}=2$ are the exponents for the Poisson and 
Wigner-Dyson distributions, respectively. 
Indeed, we have found such linear dependences characterized by different 
values of $\beta$ for varying  $E$. 
In Fig.~\ref{straight},  we show two representative dependencies for the states 
at the band centre (solid line, $\beta=1.58$) and from the tail region 
(dashed line, $\beta=1.38$).   
We should emphasize that we have used a slightly different statistic from that of 
Ref.~\cite{Cuevas_Euro}. 
This was due to the fact that the DOS in the tail regions rapidly approaches 
zero and there are a smaller number of eigenvalues (and thus level spacings) 
available for statistical analysis. 
Therefore, we have moved the range of level spacing $s$ towards smaller 
values, $s \in [1.5, 3]$, where there are more data in $P(s)$ and 
the double-log graph is still reasonably straight, 
in contrast to the higher $s$ ranges used in Ref.~\cite{Cuevas_Euro}.  
Such a change in the range of analysis resulted in slightly higher values of 
$\beta$ for the states around the band centre (see the inset in Fig.~\ref{straight}).      
We have checked our method of analysis  for the band centre 
using higher $s$ ranges  
and reproduced the published values~\cite{Cuevas_Euro} of $\beta$ for different $b$.

\begin{figure}
\includegraphics[angle=270, width=0.48\textwidth]{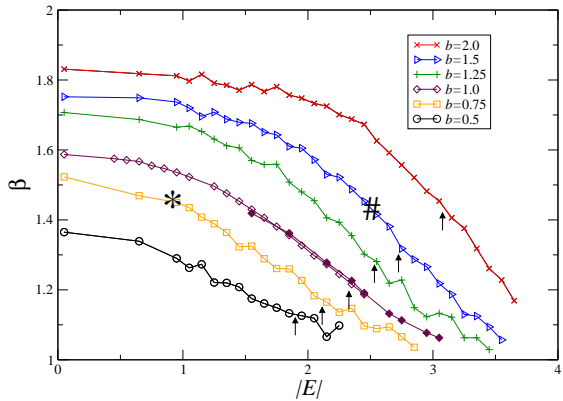}
\caption{
\label{mainresult}
The exponent $\beta$ versus the eigenvalues of the PRBM for different values 
of parameter $b$ (marked by different symbols).
The arrows indicate the eigenvalues corresponding to the 95th percentile 
of the respective DOS (see Fig.~\ref{dos}).  
Two data sets for  $b=1$ 
(open diamonds obtained for $1000\times 1000$ matrices averaged over 
$10^5$ configurations, and solid diamonds obtained 
 for $8000\times 8000$ matrices averaged over 
$420$ configurations) show the absence of a size effect.
The bin width on the eigenvalue axis was independent of eigenvalue and 
equal to 0.1 for all the models.
The symbols, * and \#, denote two pairs of parameters $(|E|, b)$ for which the level-spacing statistics have the same exponent $\beta$.
}
\end{figure}

\begin{figure}
\includegraphics[angle=270, width=0.48\textwidth]{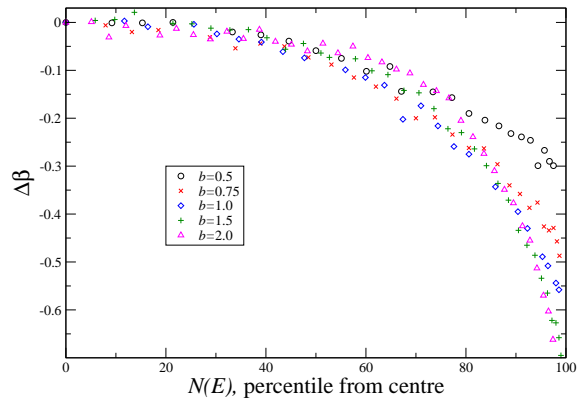}
\caption{\label{Percentile} 
The deviation, $\Delta \beta = \beta(E)-\beta(0)$, from its value at the band-centre, plotted for different values of $b$.  
The eigenvalue abscissa has been scaled to give the percentile of the density of states, with the band centre taken as zero and the far tail as 100\%.}
\end{figure}

Fig.~\ref{mainresult} shows our main result. 
In this figure, we demonstrate the possibility of mapping between different values 
of eigenvalue $E$  and parameter $b$ in terms of the same 
level-spacing statistics characterized by the exponent $\beta$. 
Indeed, as can be seen from Fig.~\ref{mainresult}, 
one value of $|E|=|E'|$ for a given $b=b'$ (e.g. the star-symbol) 
corresponds to a different value of $|E|=|E''|$
with another value of $b=b''$ (e.g. the hache symbol)  
but having the same statistics characterized by $\beta$. 
In other words,
we demonstrate that it is possible to access  different dimensionalities for 
the standard Anderson LD transition simply by analyzing different eigenvalues 
of the PRBM at fixed value of the the parameter $b$. 

The plots in Fig.~\ref{mainresult} clearly show 
a region around the band centre where 
the statistics do not change much followed by a decrease in 
$\beta$ towards the Poisson value ($\beta =1$) on approaching the edge of the band. 
The range of eigenvalues around the band centre where the value of $\beta$ is approximately 
constant depends on the value of parameter $b$. 
In order to determine the extent of this region around the band centre, 
it is illuminating to plot the change of the exponent 
$\beta$ from the band centre, $\Delta \beta = \beta(E) - \beta(0)$, versus 
the cumulative density of states, $N(E)=2\int^E_0 g(E)\text{d}E$. 
In this case, as seen in Fig.~\ref{Percentile},  
all the curves in the midgap region, say for $N(E)\alt 0.5$, 
 approximately  collapse to a single curve. 
This justifies an analysis of the level-spacing statistics around the band centre 
for less than approximately 50$\%$ of states with an absolute error in $\beta$ less than $0.1$. 

The data presented in Fig.~\ref{mainresult} were obtained for $8000\times 8000$ matrices 
by averaging over 420 realizations for all values of $b$ except $b=2$, for which 620 
realizations were used to obtain better statistics in the far tail. 
The absence of size effects has been checked for the case of $b=1$ by performing 
additional analyses for $1000\times 1000$ matrices (compare open and solid squares 
in   Fig.~\ref{mainresult}).

\begin{figure}
\includegraphics[angle=270, width=0.48\textwidth]{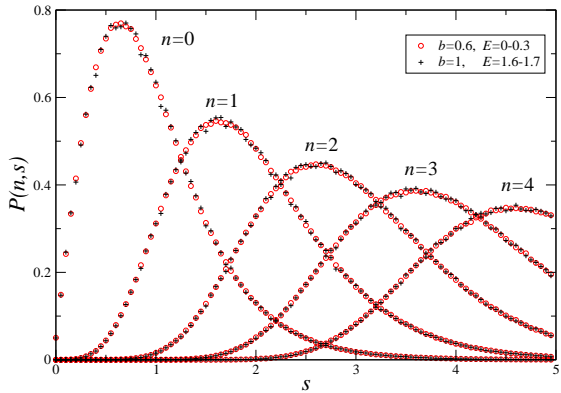}
\caption{
\label{P_n_s}
The $n$-level spacing probability distribution functions $P(n,s)$ ($n=0, ..., 4$) 
for two sets of parameters $(b'=0.6, |E'|\in[0,0.3])$ and $(b''=1, |E''|\in[1.6,1.7])$, 
both characterized by the same value of $\beta\simeq 1.41$. 
For generating these graphs, $1000\times 1000$ matrices averaged over 
$25000$ ($b'=0.6$) and $10^5$ ($b''=1$) realizations were used.  
}
\end{figure}

In order to verify the equivalence between different values of 
$(b,|E|)$ in terms of equivalent level-spacing statistics, found by analysis to have 
the same value of the exponent $\beta$ for the adjacent level-spacing 
probability distribution $P_{\text{c}}(s)$, given by Eq.~(\ref{e1}), we have 
calculated in addition the $n$-level distribution functions, $P_{\text{c}}(n,s)$\cite{Mehta:book}. 
These functions are the probability distribution functions for the level spacing $s$ 
between two levels separated by $n$ other levels, with $P_{\text{c}}(n=0,s)\equiv 
P_{\text{c}}(s)$.   
Fig.~\ref{P_n_s} shows the equivalence between all five  $n$-level distribution functions 
$(n=0,\ldots, 4)$ for two representative pairs of parameters, $(b',|E'|)$ and $(b'',|E''|)$
characterized by the same value of $\beta(\simeq 1.41)$. 
This means that the statistical properties of the critical eigenvalues, which 
can be characterized by different functions, e.g. the 
two-point correlation function \cite{Braun_95}, 
$R_{\text{c}}(s)=\delta(s)-1+\sum_{n=1}^{\infty}P_{\text{c}}(n,s)$, are the same 
for   $(b',|E'|)$ and $(b'',|E''|)$. 
 
In conclusion, we have provided numerical evidence for the equivalence of 
statistical properties of critical eigenstates for different values of both the 
strength parameter $b$ and eigenvalues of power-law random banded matrices (PRBMs). 
This justifies the use of an alternative way of exploring different dimensionalities 
for the standard Anderson transition. 
Instead of the usual way of analyzing the level-spacing statistics around the band centre 
for different values of $b$, we show that it is possible to access 
the same statistics by  moving through the band of eigenvalues at fixed value of $b$. 
Also, we specify the range of the eigenvalues around the band centre where the level-spacing 
statistics do not change significantly, and thus justify the standard method of 
statistical analysis of eigenvalues of the PRBM by varying $b$.

CJP would like to thank the Engineering and Physical Sciences Research Council for financial support.

\bibliography{archive_cjp2}


\end{document}